\documentclass[final]{IOS-Book-Article}
\pdfoutput=1

\usepackage{graphicx}
\graphicspath{ {figures/} }
\usepackage{amsmath}
\usepackage{url}
\usepackage{hyperref}
\usepackage{multirow}
\usepackage{booktabs}
\usepackage{tabularx}
\usepackage{multirow}
\usepackage[caption=false,font=footnotesize]{subfig}

\usepackage{mathptmx}
\def\hb{\hbox to 10.7 cm{}}

\usepackage{makecell}
\usepackage{gensymb}

\begin{document}

\pagestyle{headings}
\def\thepage{}

\begin{frontmatter}              


\title{A Methodology for Obtaining Objective Measurements of Population Obesogenic Behaviors in Relation to the Environment}

\markboth{}{October 2019\hb}

\author[A]{\fnms{Christos} \snm{Diou}%
\thanks{Corresponding Author: Christos Diou, Tel: +30-2310-994376, E-mail: \url{diou [at]
auth.gr}.\\
  The final publication is available at IOS Press through \url{http://dx.doi.org/10.3233/SJI-190537}}},
\author[A]{\fnms{Ioannis} \snm{Sarafis}},
\author[A]{\fnms{Vasileios} \snm{Papapanagiotou}},
\author[B]{\fnms{Ioannis} \snm{Ioakimidis}}
and
\author[A]{\fnms{Anastasios} \snm{Delopoulos}}
\runningauthor{C. Diou et al.}

\address[A]{Department of Electrical and Computer Engineering, Aristotle University of Thessaloniki, Greece}
\address[B]{Department of Biosciences and Nutrition, Karolinska Institutet, Stockholm, Sweden}

\begin{abstract}
The way we eat and what we eat, the way we move and the way we sleep
significantly impact the risk of becoming obese. These aspects
of behavior decompose into several personal behavioral elements
including our food choices, eating place preferences, transportation
choices, sleeping periods and duration etc. Most of these elements are
highly correlated in a causal way with the conditions of our local
urban, social, regulatory and economic environment. To this end, the
H2020 project ``BigO: Big Data Against Childhood Obesity''
(http://bigoprogram.eu) aims to create new sources of evidence
together with exploration tools, assisting the Public Health
Authorities in their effort to tackle childhood obesity. In this
paper, we present the technology-based methodology that has been developed
in the context of BigO in order to: (a) objectively monitor a matrix
of a population's obesogenic behavioral elements using commonly available
wearable sensors (accelerometers, gyroscopes, GPS), embedded in smart
phones and smart watches; (b) acquire information for the environment
from open and online data sources; (c) provide aggregation mechanisms
to correlate the population behaviors with the environmental
characteristics; (d) ensure the privacy protection of the
participating individuals; and (e) quantify the quality of the
collected big data.

\end{abstract}

\begin{keyword}
big data \sep wearables \sep population behavior \sep objective measurements \sep obesity
\end{keyword}
\end{frontmatter}

\section{Introduction}
\label{sec:introduction}
Obesity is highly prevalent among children and adolescents in Europe, with a particularly
high rate in children of families with low socioeconomic status \cite{Bammann2013}. On
average, obesity affects one in every three children aged six to nine years in Europe
\cite{Wijnhoven2014}, with no other chronic disease reaching such high prevalence in the
school-aged population. Children who are obese are more likely to stay obese into
adulthood, which puts them at increased risk for non-communicable diseases (NCDs), such as
type-2 diabetes and cardiovascular disease. This fact, combined with the slow but
continuous increase in the obesity prevalence in the last forty years
\cite{EUActionPlan2014} jeopardizes the sustainability of our healthcare systems.

The World Health Organization’s (WHO) Commission on Ending Childhood Obesity has recently
released a comprehensive report \cite{WHOReport} outlining a high-level set of
recommendations to tackle the childhood obesity epidemic, grouped into 6 broad categories:
(i) promoting healthy foods, (ii) promoting physical activity (PA), (iii) preconception
and pregnancy care, (iv) early childhood, (v) school-aged children, and (vi) weight
management for overweight and obese children. Cross-cutting through all recommendations is
the need for ``robust monitoring and accountability systems'', which ``are vital in
providing data for policy development and in offering evidence of the impact and
effectiveness of interventions.'' Furthermore, the report recognizes that successful
measures should address the entire obesogenic environment. There are several difficulties
towards implementing these recommendations. Individuals and their behavioral choices are
situated within and influenced by their broader social and environmental context \cite{Lyn2013},
which consists of a complex array of local external factors \cite{Lobstein2015}, like community,
demographic and socioeconomic characteristics. Measures that combine multiple strategies
that modify the obesogenic environment may improve the dietary and sleeping habits,
increase physical activity and reduce sedentary behaviors. Such interventions can be successful
\cite{Wang2013}, if they are evidence-based and context-specific \cite{DeBour2015}.

Smartphones, portable or wearable sensors, open data and Internet of Things technologies
can act as an enabler for \emph{objectively} measuring the information required to study
the obesogenic behaviors of the population in relation to their local environment, and to
effectively design appropriate policies. Although the use of sensors on mobile phones and
wearables is now common for lifestyle and sports applications, using them for developing
evidence-based policies against childhood obesity is not straightforward. Issues to be
resolved include unobtrusive data collection, protection of the privacy and anonymity of
participants, selection of variables to be calculated, data aggregation as well as data
quality.

In this paper we present an overview of the data acquisition and aggregation methodology
that is currently being developed in the BigO project \cite{bigosite} (12/2016 -
12/2020). The proposed methodology aims to address the above issues and provide
practical solutions for extracting statistical evidence from big data collected from
multiple heterogeneous data sources in an uncontrolled manner.

The proposed methodology includes steps for (i) data acquisition for objective measurement
of individual behavioral indicators, (ii) measurement of environment factors that are
relevant to childhood obesity, (iii) mechanisms for aggregating individual data to
measurements describing the behavior of the population (to support analysis of behaviors
with respect to the environment), (iv) mechanisms for controlling the level of privacy
protection of individuals and (v) preliminary steps for quantifying data quality and
addressing quality issues in the data analysis processes.

These elements of the BigO methodology are general, in the sense that can easily be
adapted for use in other domains beyond childhood obesity. In addition, they are aligned
with the principles of Trusted Smart Surveys, as they are outlined in
\cite{Ricciato2019}. Specifically, both propose the use of passively collected sensor data
to extract objectively collected measurements in order to augment data collection from
survey participants. Furthermore, both propose mechanisms for strong data and privacy
protection. For example, Secure Multi-party Computation approaches, such as
\cite{Archer2018} and \cite{Zyskind2015}, have been suggested for Trusted Smart Surveys,
while the privacy controlling mechanism of BigO is outlined in Section
\ref{sec:aggregation}. Finally, both Trusted Smart Surveys and BigO rely on voluntary
citizen participation through the concepts of ``Citizen Statistics'' and ``Citizen
Science'', respectively. Based on these observations, we can argue that in terms of
methodology, the proposed approach is aligned with the Trusted Smart Statistics concept
\cite{BucharestMemorandum2018}. They differ, however, in terms of their objective
(official statistics vs scientific research).

The structure of this paper is as follows. Section \ref{sec:information_model} outlines
the information model of the proposed methodology, from raw data acquisition to data
analysis. Then, Section \ref{sec:big_data_evidence} goes deeper in the description of the
types of behaviors and environmental factors that are considered for the problem of
childhood obesity. Section \ref{sec:aggregation} focuses on the mechanisms for aggregating
individual data to population behavior, while Section \ref{sec:quality} describes the
mechanism for quantifying data quality according to the proposed methodology. Section
\ref{sec:lessons_learned} discusses the challenges and lessons learned during the
development of BigO, which are relevant to Trusted Smart Statistics and especially Trusted
Smart Surveys, and will hopefully be useful to developers of such systems. Finally,
Section \ref{sec:conclusions} concludes this work and provides directions for future
research.

\section{Information model and analysis methodology}
\label{sec:information_model}

The primary goal of the examined methodology is to create new sources of evidence together
with exploration tools for the Public Health Authorities to assist in their effort in
developing policies against childhood obesity. Information in BigO can be grouped into
three layers: (a) the raw data from sensors and external sources; (b) the quantification
of behavioral and environmental characteristics; and (c) models and analytics. An
illustration of the BigO information model is given in Fig. \ref{fig:information_model}.

Overall, the methodology of BigO relies on the following functionalities:

\begin{figure}[!t]
\centering
\includegraphics[width=.9\textwidth]{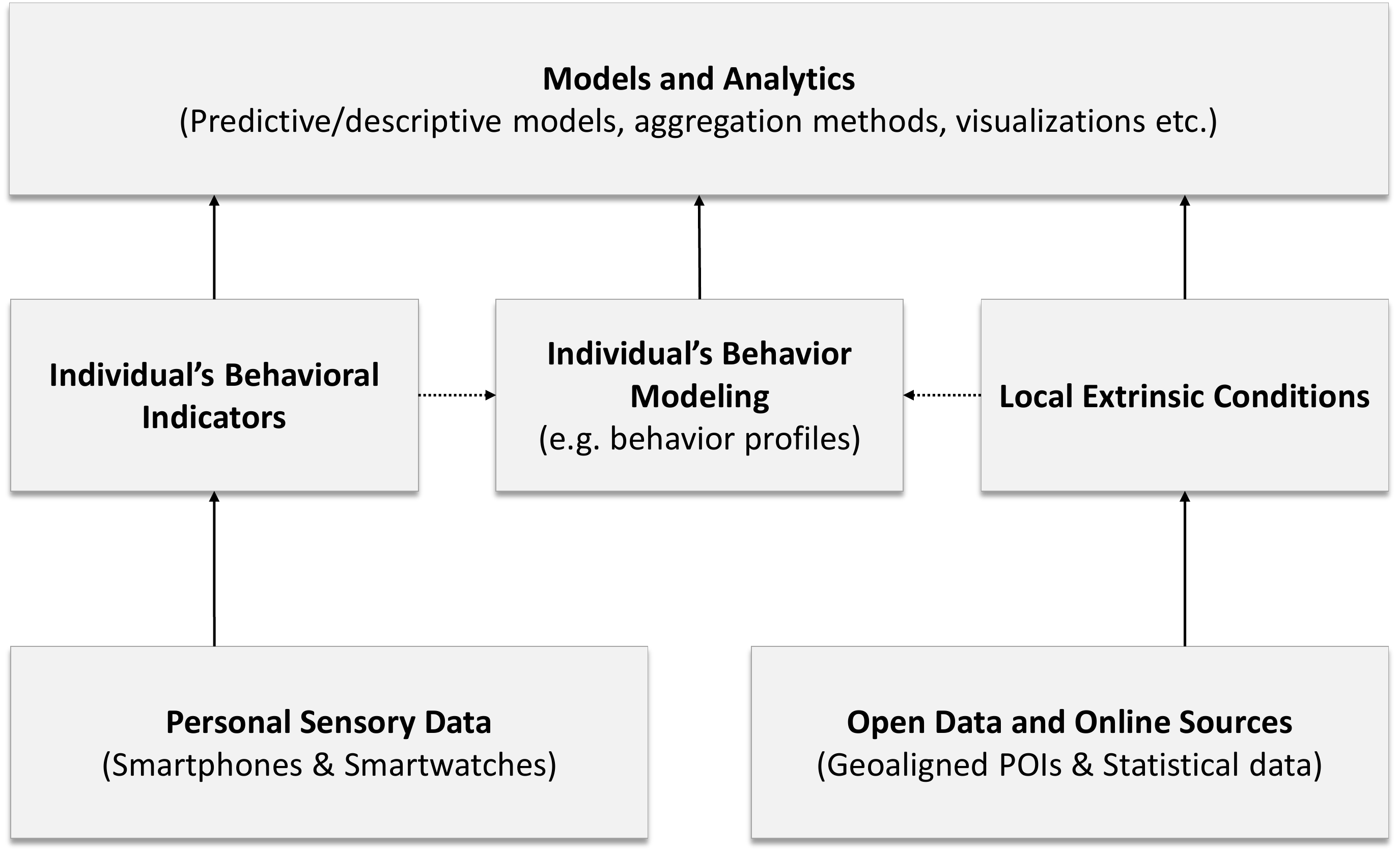}
\caption{BigO's layers of information}
\label{fig:information_model}
\end{figure}

\begin{enumerate}

\item The collection of Big Data associated with the individuals' behavioral patterns
  (e.g., based on accelerometry and geolocation), using different technologies (smart
  phones, smart watches and wristbands). The collected sensor data are processed to
  produce \emph{behavioral indicators}, which quantify behaviors known to be associated
  with obesity, such as eating habits and diet, physical activity and sleep.

\item The collection of Big Data about characteristics of the environment which may affect
  the local population behavior and, eventually, contribute to the development of
  unhealthy lifestyle habits. This type of data is collected from multiple online and
  publicly available sources, such as official statistics, maps, registries and Geographic
  Information Systems (GIS). The collected data are processed to calculate the \emph{Local
    Extrinsic Conditions} (LECs) which represent the local context in terms of the urban
  landscape, school programs, local policies, socioeconomic factors and food marketing.

\item The creation of comprehensive models of the obesity prevalence dependence matrix,
  through the association of the LECs with the obesogenic behavioral patterns of the
  population. Note that the behavioral indicators and LEC data --- instead of the original
  raw data --- are used as input to the models. These models are the basis for providing
  data-driven decision support to public health authorities, policy makers and
  clinicians. Specifically, the targeted models are used to
  \begin{enumerate}
    \item Identify the most important obesogenic factors of the local
      environment. Although it is in general known what are the main conditions of the
      urban, the social, the regulatory and legal environment that negatively affect the
      obesogenic behavior, the examined methodology aims at identifying those that are
      prominent at a local level.
    \item Simulate the effect of interventions to the obesogenic behaviors. Local
      authorities will have indication of the effectiveness of their counter-obesity
      measures before their actual implementation.
  \end{enumerate}

\item The visualization of the acquired data and their relations is
  also part of the functionalities that facilitate the exploration of
  populations behaviors versus the local environment conditions.
\end{enumerate}

\section{Using big data for collecting evidence}
\label{sec:big_data_evidence}

This section describes in more detail the raw data acquisition and processing methodology
of BigO (layers 1 and 2 of the information model shown in Fig.
\ref{fig:information_model}). Sections \ref{sec:raw_data} and \ref{sec:online_data}
describe the data sources for behavior and local environment, whereas Sections
\ref{sec:behaviour_indicators} and \ref{sec:lecs} describe the behavioral indicators
and LECs, respectively. Section \ref{sec:behavioural_profiles} demonstrates a
mechanism for describing the temporal characteristics of an individual's behavior. Finally
Section \ref{sec:bias} briefly discusses the process for participant
selection in BigO.

\subsection{Data collection}

\subsubsection{Sensor data}
\label{sec:raw_data}
The first source of raw data used in BigO is \emph{Personal Sensory Data} acquired by
smart phones and commercial smart watches. These are raw sensory data that are collected
via a portable/wearable device, pertain to the individual that is using the device (with
the single exception of food advertisement photographs), and are related to the behaviors
that are of interest to BigO (what one eats, how one eats, how one moves, etc.). BigO
relies on two types of sensor-enabled devices: (a) smart phones with Android or iOS and
(b) smart watches with wearable OS (e.g. Wear OS).

The sensory data collected via the smartphones are inertial measurement unit data (IMU),
location data and photographs. Other types of sensors are available which, however, are
currently not used. Smartwatches collect the same types of data with the exception of
photographs. Photographs are captured whenever users decide to submit pictures of their
meals or food-related advertisements.

Acquisition of the raw data is constrained by battery consumption limitations that
currently prohibit continuous acquisition at high sampling frequency. Typically, all
sensor signals available on the device are recorded on the slowest sampling frequency the
operating system allows, which is enough for the calculation of most behavioral indicators
in BigO. For most devices, the sampling rate for IMU sensors is between $5$ an $25$ $Hz$,
while for GPS location data sampling rate is reduced to one sample per minute.

There are certain types of indicators that can only be calculated using raw data at high
sampling frequency. For example, our algorithms for automatic extraction of in-meal
behavior indicators (e.g. bite detection) using smart watches
\cite{Kyritsis2019, Kyritsis2018, Papadopoulos2018} require triaxial accelerometer and
gyroscope signals with sampling frequency over $60 Hz$. Integrating such indicators
requires data collection at high sampling frequency for relatively small time periods
during the day (e.g. only during meals).

\paragraph{Data acquisition mechanisms for battery preservation}
\label{sec:doze}
To further preserve battery, BigO's raw data acquisition software is aware of OS-level
optimizations (such as the "Doze" mode for Android phones \cite{Doze}). Battery
optimizations essentially set the phone (or watch) CPU and sensors to stand by when it is
not used. The criteria for what counts as usage vary slightly across different operating
systems and device vendors. In practice, sensor recording stops when the device is left
stationary with screen switch off for more than a few minutes. The device resumes
recording as soon as there is some user input or significant motion. This fits the
use-case of BigO very well since the criteria for suspending sensor recording directly
imply that the device is not used, or that the individual's physical activity is very low
(and therefore no need to record sensor data).

\subsubsection{Open data and online sources}
\label{sec:online_data}
The second source of data used in BigO are the \emph{Environment Data} retrieved from
online data or public data providers: geo-aligned Points of Interest and statistical data.

\paragraph{Geo-aligned Points of Interest}
A number of online providers, such as OpenStreetMap, Google Maps, Foursquare, and Bing
Maps, provide access to the metadata of Public Points of Interests (POIs). Among the
different types of POIs, in BigO we are interested in POIs that refer to:
\begin{itemize}
\item Food, such as restaurants, fast food outlets, grocery stores,
  supermarkets.
\item Physical activity, such as gyms, pools, sports.
\item Transportation, such as bus stops, metro stations.
\item Other types of facilities related to child behavior, such as parks and indoor
  recreation facilities.
\end{itemize}
Additional metadata may characterize the usage of the POIs (e.g., timetables,
transportation routes). We have adopted a common coding scheme
and heuristic rules to map the characterizations of each data source/provider to an
internal taxonomy.

\paragraph{Demographic, Social and Financial Statistical Data}
Socioeconomic indicators such as the average income, unemployment rates, type of employment,
educational level are collected from published archives of Eurostat and the National
Statistical Authorities. This type of information is directly related to the LECs and
heavily influence our aetiology models.

Other data on behavior, habits, and lifestyle of the population and on obesity prevalence
have been collected and are openly available from WHO and initiatives like the Childhood
Obesity Surveillance Initiative (COSI) \cite{COSI}.

\paragraph{Complementary sources for environment data}
Two important limitations of the statistical data sources are (a) the
relatively coarse spatial resolution of the statistical data (NUTS 1,
2 and rarely 3) and (b) the relatively low temporal resolution limited
by the census periodicity. Two complementary sources of similar
information are being currently explored:

\begin{figure}[!t]
\centering
\subfloat[Pylaia]{
  \includegraphics[width=0.49\linewidth]{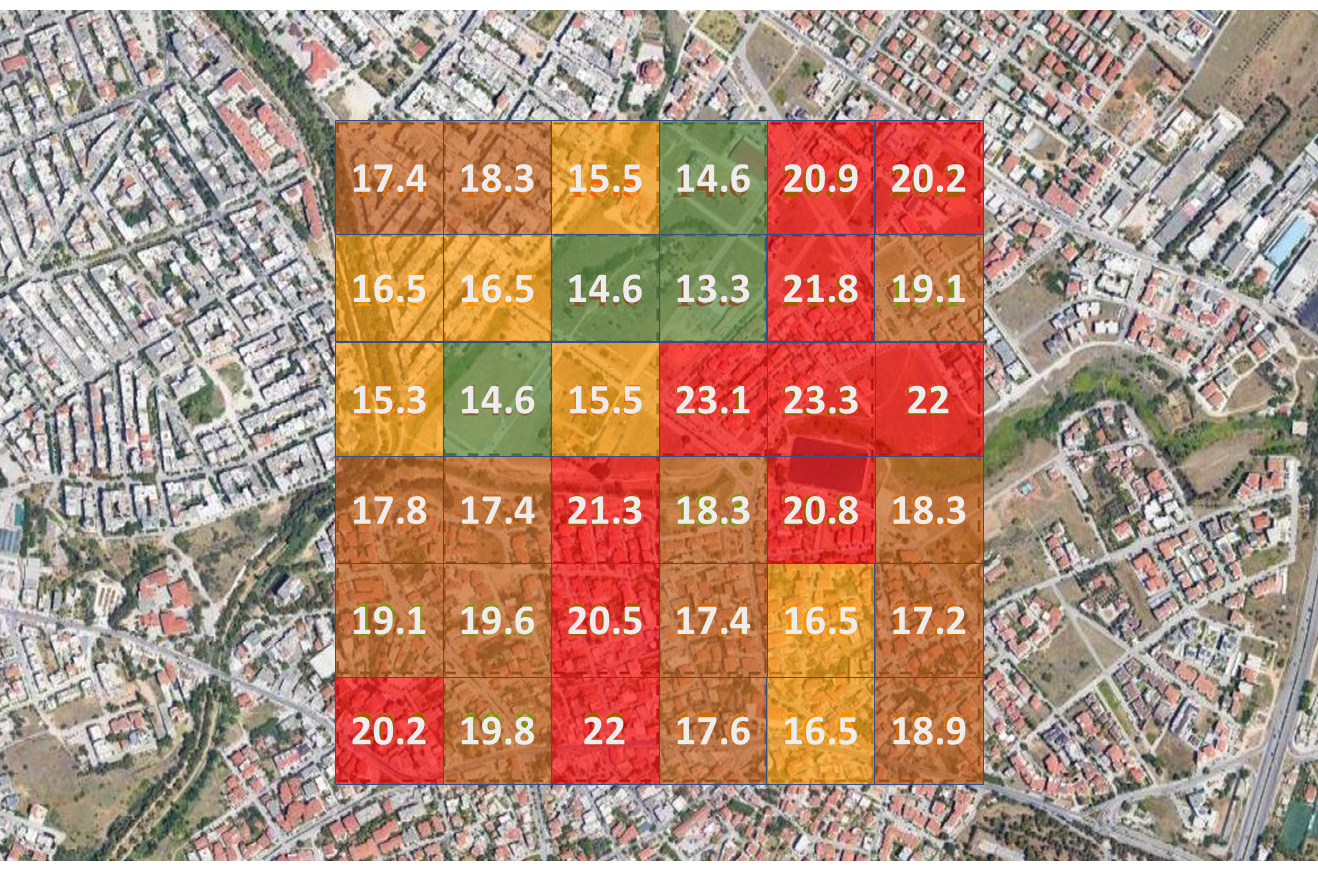}
}
\subfloat[Panorama]{
  \includegraphics[width=0.49\linewidth]{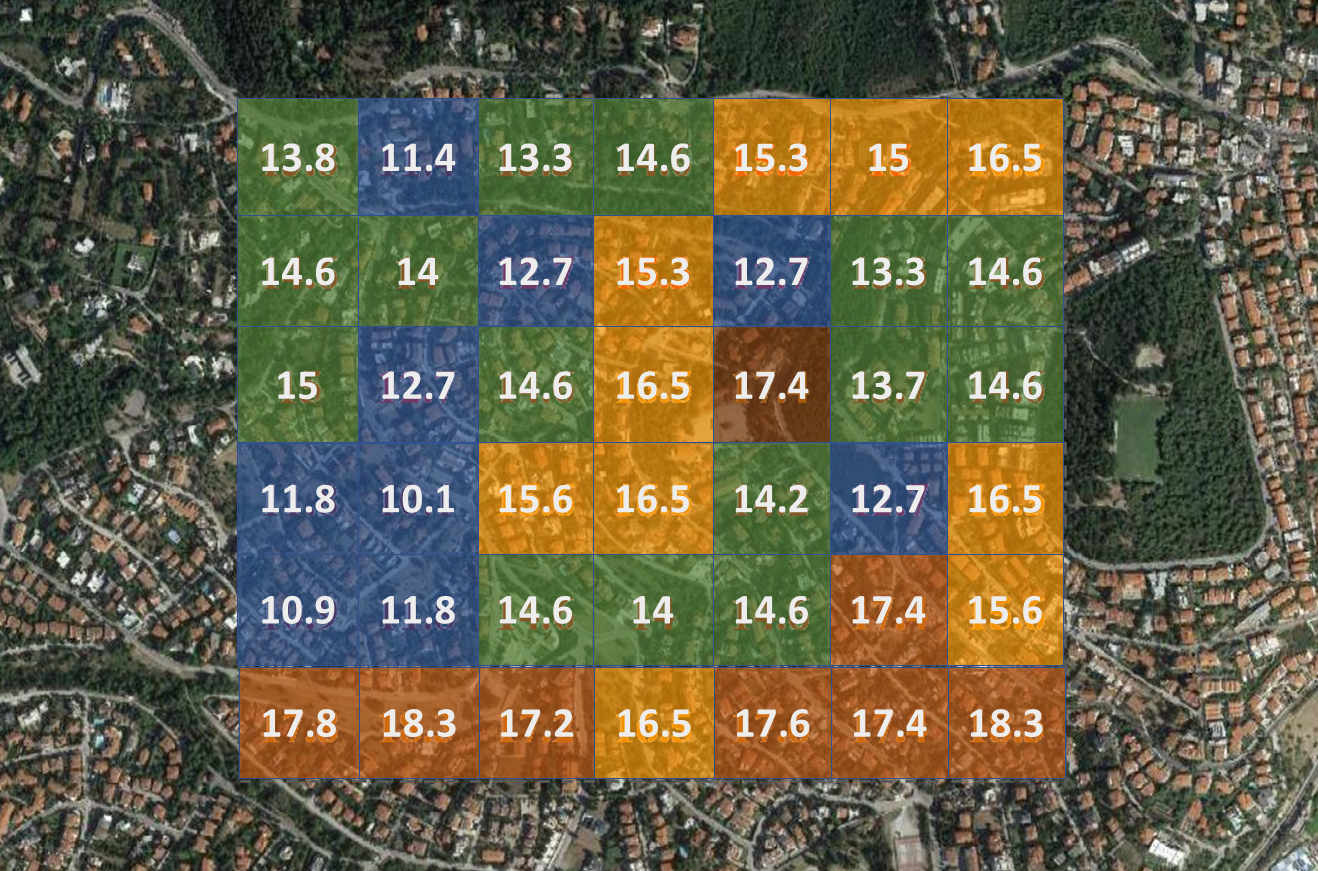}
}
\caption{Unemployment rate in blocks of two areas, inside the same
  Greek municipality. Orange and red values indicate high, while blue
  and green values indicate low unemployment rates as estimated by the
  analysis of car images appearing on Google Street View
  \cite{Diou2018}.}
\label{fig:unemployment_map}
\end{figure}

\begin{enumerate}

\item The microdata repositories that are kept by the statistical
  authorities.
\item Inference of statistics of interest from the analysis of publicly available data. A
  very promising illustration of this option is our work in \cite{Diou2018} that attempts
  to predict unemployment rate at a fine resolution by applying deep learning and image
  processing techniques to Google street view images. An example of the method for two
  municipalities in Greece is given in Fig. \ref{fig:unemployment_map}. In this example,
  the true unemployment rate is estimated using a linear model and a surrogate variable
  calculated automatically from the parked cars in Google street view images ($R^2= 0.76$,
  correlation coefficient is $0.874$). Similar results were also achieved for other
  statistical variables related to education level and occupational prestige.

\end{enumerate}

\subsection{Behavioral indicators and LECs}

\subsubsection{Behavioral indicators}
\label{sec:behaviour_indicators}

In BigO, the \emph{behavioral indicators} are measurable quantities that provide
information for the behavior of an individual. Overall, the behavioral indicators are
measures that describe an individual's behavior on diet (what you eat), eating behavior
(how you eat), physical activity (how you move) and sleep.

In addition to the categorization in terms of the type of behavioral measurement,
indicators differ in terms of the way they are computed. According to this viewpoint,
three types of indicators are identified in BigO:

\begin{enumerate}
\item \emph{Self-reported indicators}. These can be computed directly from the
  individual's self-reports. Their drawback is that they depend on the user
  compliance and reporting accuracy; thus, they tend to be unreliable.
\item \emph{Base indicators}. These are indicators that are computed directly by
  processing the Personal Sensory Data. Their advantages are that they are calculated
  automatically, they provide objective measurements of behavior and they do not require
  any effort by the individual. Table \ref{table:behavioural_indicators_base} shows
  examples of base indicators in BigO. (Note that the ``diet'' indicators are also
  collected as self-reported indicators.)
\item \emph{Derived indicators}. These are calculated from the base indicators and may
  also leverage self-reported data. Table \ref{table:behavioural_indicators_derived} shows
  examples of derived indicators.
\end{enumerate}
\begin{table}[!t]
  \begin{minipage}{\textwidth}
  \caption{Indicative list of ``base'' behavioral indicators}
  \renewcommand{\arraystretch}{1.1}
  \label{table:behavioural_indicators_base}
  \begin{tabularx}{\linewidth}{Xll}
    \toprule
    \textbf{Name} & \textbf{Units} & \textbf{Sensors}
      \footnote{\textbf{L}: Location-related sensors, such as GPS,
        magnetometer. Either on the mobile phone or in wristband/smartwatch,
        \textbf{A}: Activity-related indicators, such as accelerometer,
        gyroscope. Either on the mobile phone or in wristband/smartwatch, \textbf{P}:
        Smartphone camera, \textbf{U}: User self-reports
      } \\
    \midrule
    \textbf{Diet Indicators} & & \\
    Eating fast food & Occurrence & L, P, U \\
    Eating dinner outside of the home & Occurrence & L, P, U \\
    Eating at home & Occurrence & L, P, U \\
    Meal type (breakfast, lunch, dinner, snack) & Categorical & L, P, U \\
    \midrule
    \textbf{Physical Activity Indicators} - Calculated at minute intervals\\ 
    Energy expenditure & MET\footnote{MET: Metabolic Equivalent of Task} & A \\
    Activity type & Categorical & A \\
    Activity intensity & Categorical & A \\
    Activity level & Categorical & A \\
    Activity counts\cite{Tryon1996} & Counts/Minute & A\\
    \midrule
    \textbf{Sleep Indicators}\\
    Hours of sleep per night & Hours & A \\
    Sleep/wake-up times per night & Timestamp & A \\
    Interruptions of sleep & Number & A \\
    Movement during sleep & Categorical & A \\
    \bottomrule
  \end{tabularx}
  \end{minipage}
\end{table}

 \begin{table}[!b]
  \begin{minipage}{\textwidth}
  \caption{Indicative list of ``derived'' behavioral indicators}
  \renewcommand{\arraystretch}{1.1}
  \label{table:behavioural_indicators_derived}
  \begin{tabularx}{\linewidth}{Xl}
    \toprule
    \textbf{Name} & \textbf{Units} \\
    \midrule
    \textbf{Diet Indicators} & \\
    Fast-food eating frequency & Times/week  \\
    Adherence to eating schedule & Minutes (Standard deviation) \\
    Food type eating frequency & Times/week \\
    Meal type frequency & Times/Week \\
    \midrule
    \textbf{Physical Activity Indicators} - Calculated daily or weekly & \\
    Walking/cycling to/from school & Times/week  \\
    Minutes of active commute to school & Minutes/day \\
    Exercise frequency & Times/week \\
    Minutes of sedentary behaviors after school & Minutes/day \\
    Distribution of physical activity at school & Minutes per activity \\
    Distribution of physical activity after school & Minutes per activity \\
    \midrule
    \textbf{Sleep Indicators}\\
    Average hours of sleep per night & Hours \\
    Average number of interruptions of sleep & Number \\
    \bottomrule
  \end{tabularx}
  \end{minipage}
\end{table}

\subsubsection{Local Extrinsic Conditions (LECs)}
\label{sec:lecs}
Local Extrinsic Conditions (LECs) are measurements of the environment extracted by
processing the Environment Data from open and online sources. LECs quantify the
characteristics of the environment which can affect an individual's behavior, including
urban landscape, school programs and policies, socioeconomic factors as well as food
marketing. Table \ref{table:lecs} shows an indicative list of LECs calculated in BigO.

\begin{table}[!b]
  \begin{minipage}{\textwidth}
  \caption{Indicative list of Local Extrinsic Conditions (LECs)}
  \renewcommand{\arraystretch}{1.1}
  \label{table:lecs}
  \begin{tabularx}{\linewidth}{Xll}
    \toprule
    \textbf{Name} & \textbf{Units} & \textbf{Sensors}
      \footnote{\textbf{L}: Location-related sensors, such as GPS,
        magnetometer. Either on the mobile phone or in wristband/smartwatch,
        \textbf{E}: External sources (e.g. Google maps),
        \textbf{M}: Media monitoring reports
      } \\
    \midrule
    \textbf{Urban Environment} & & \\
    Availability of supermarkets and grocery stores & Yes/No, count \& location & E, L \\
    Availability of restaurants and food outlets & Yes/No, count \& location & E, L \\
    Availability of take-away restaurants & Yes/No, count \& location & E, L \\
    Availability of cafes/bars & Yes/No, count \& location & E, L \\
    Availability of wine/liquor stores & Yes/No, count \& location & E, L \\
    Availability of public parks & Yes/No, count \& location & E, L \\
    Availability of indoor recreational facilities & Yes/No, count \& location & E, L \\
    Availability of outdoor recreational facilities & Yes/No, count \& location & E, L \\
    Open spaces in neighborhood & Percentage/Categorical & E, L \\
    Density of food outlets & Number/km$^2$ & E, L \\
    Number of food outlets within a 100m/1000m radius & Number & E, L \\
    Number of recreational facilities within a 100m/1000m radius & Number & E, L \\
    Density of food outlets & Number/km$^2$ & E, L \\
    Density of recreational facilities & Number/km$^2$ & E, L \\
    Distribution of recreational facility type & Percentage/Categorical & E, L \\
    \midrule
    \textbf{School Environment} & & \\
    School exercise programs & Times/week, Duration & E \\
    School meals/breaks & Number, Duration & E \\
    School hours & Start/end timestamps & E \\
    \midrule
    \textbf{Socioeconomic Environment} & & \\
    Average income in neighborhood & EUR(SEK)/person/year & E \\
    Education level statistics & Education level distribution & E \\
    Unemployment rates & Percentage & E \\
    \midrule
    \textbf{Food marketing} & & \\
    Exposure to food advertising from TV & Categorical, Ads/day & M \\
    Exposure to food advertising in urban environment & Number of food ads in area & M, U \\
    Food advertising at specific times & Series of timestamps & M, U \\
    \bottomrule
  \end{tabularx}
  \end{minipage}
\end{table}

\subsection{Behavioral profiles}
\label{sec:behavioural_profiles}

The base indicators are the first step for quantifying the individual's
behavior. Subsequently, the derived indicators provide some higher level information about
the individual. However, indicators cannot express all aspects, especially the ones that
refer to temporal characteristics of behavior. For example, consider the following cases:
\begin{itemize}
  \item After school, does the individual go to home or not? 
  \item If not, what types of POIs does the individual visit?
  \item When is it most likely to visit a fast food restaurant or take
    away outlet?
  \item What type of POIs precede a visit to such facilities for the
    individual?
\end{itemize}

\begin{figure}[!t]
\includegraphics[width=.6\textwidth]{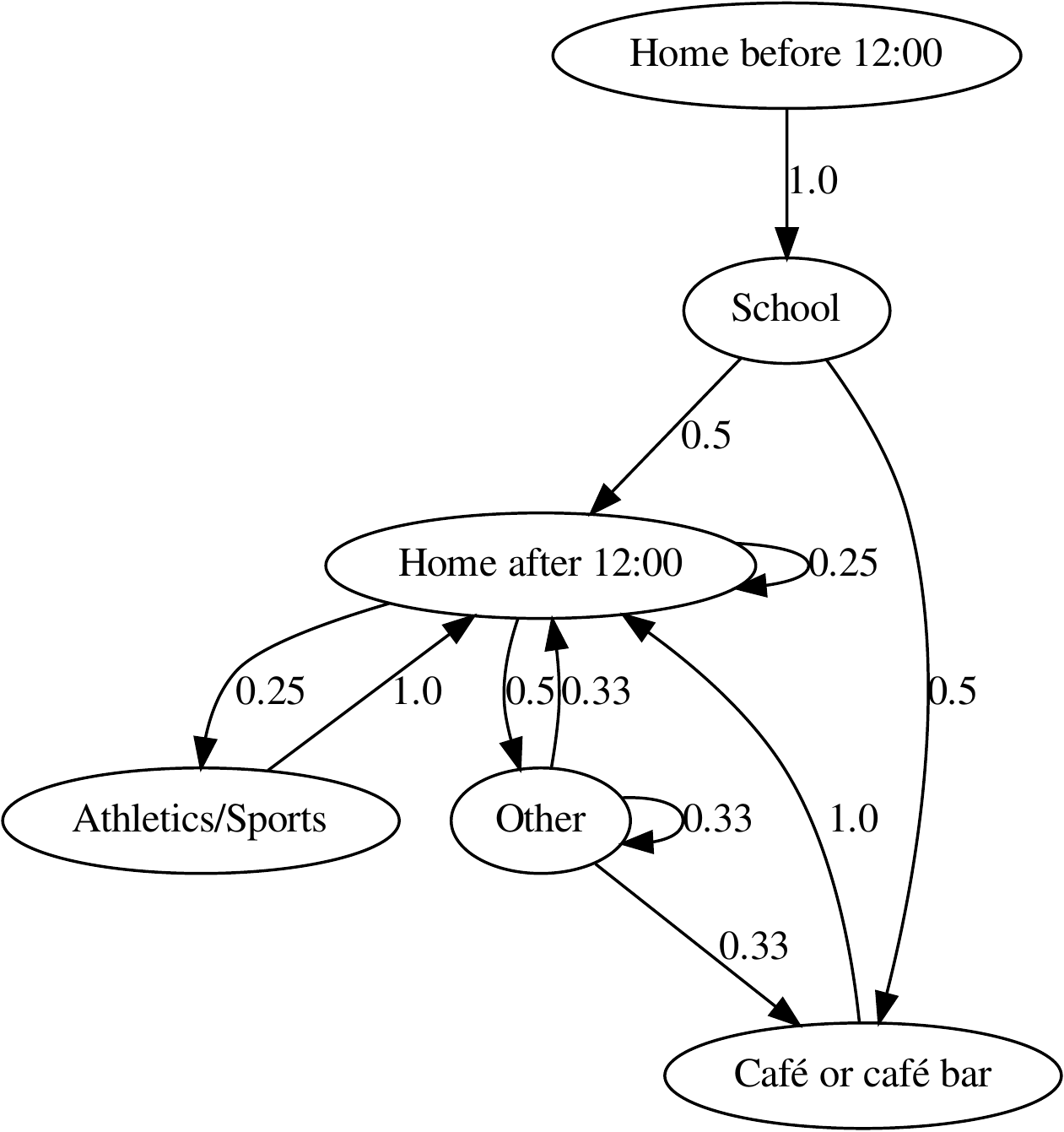}
\caption{Example graph visualization of a behavior profile
  \cite{Sarafis2019}. The profile was extracted using the timelines of
  $10$ school days for a student that participated in a BigO pilot.}
\label{fig:behavior_profile}
\end{figure}

For this purpose, BigO has developed mechanisms to systematically model the temporal
characteristics of behavior (i.e. the individual's daily habits), which are known to be
associated with the risk of developing obesity. The most prominent mechanism is the
\emph{behavior profiles} \cite{Sarafis2019}.

Briefly, a behavior profile for an individual is calculated through the following steps:
\begin{enumerate}
  \item For each individual we identify the visited POIs by executing a clustering
    algorithm. For example, we can use the DBSCAN variant of Luo \emph{et al.}
    \cite{Luo2017}, which is tailored for geospatial trajectories. The POIs are then
    transformed to reflect their type (e.g. school, fast food or take away, athletics and
    sports, public parks, etc.)  using online data sources (such as, Google maps and
    Foursquare). The more important element is that the actual coordinates of the POIs are
    discarded in the next steps. This way we can ensure high level of privacy protection
    since the actual location coordinates are never used.

  \item The timeline is constructed as a sequence of ``stop'' and ``move'' events. Each
    event contains the recorded sensory data and extracted base indicators, the
    timestamps, the POI type for ``stop'' events and the origin and destination POI types
    for ``move'' events. In addition, ``move'' events contain the travel distance and the
    transportation mode (e.g. vehicle or walking).

  \item Using the available timelines of an individual we calculate two behavior profiles,
    one for school days and one for non-school days. A behavior profile consists of three
    parts:

    \begin{enumerate}
      \item A \emph{transition graph} that captures the individual's mobility patterns. It
        is calculated using the frequency between the origin and destination POIs of the
        ``move'' events across all timelines of the specific type. It is based on the
        assumption that the timelines can be modeled by a first order Markov chain for the
        mobility patterns of the individual \cite{Jahromi2016}. An edge of the transition
        graph from POI type $i$ to POI type $j$ has the transition probability:

        \begin{equation*}
          P_{ij} = Pr\{ \text{Transition from POI type } i \text{ to type } j ~|~
          \text{Individual is at POI type } i\}
        \end{equation*}

        Fig. \ref{fig:behavior_profile} shows an example graph visualization of a
        behavior profile calculated for a student that participated in a BigO pilot using
        the timelines of $10$ school days.

      \item The \emph{transition metadata} for each edge of the graph with $P_{ij} >
        0$. They describe the transportation mode preferences (e.g. vehicle, walking) as a
        probability mass function. In addition, they model for each transportation mode:
        the travel distance, the travel duration and the recorded physical activity
        indicators during the transition between the POIs. The metadata variables can be
        modeled using distributions or their average values.

      \item The \emph{POI metadata} for each node of the graph. They describe how the
        individual behaves during a visit at each POI type (e.g. number of meals, physical
        activity indicators) and can be modeled using distributions or their average
        values.
    \end{enumerate}
\end{enumerate}

More information for the calculation of behavior profiles and detailed examples from
real-world data can be found in \cite{Sarafis2019}.

\subsection{A note on participant selection and bias}
\label{sec:bias}
The presented methodology focuses on how data is collected from each individual
participant and each geographical region, independently of how these have been
selected. For the statistical analysis the sampling mechanism is of high importance,
especially if the objective is to compute statistics about the population.

In BigO, children participate through their schools in the context of class
activities, and under the supervision of their teachers. Participation is voluntary and
there are no exclusion criteria. High school children participate using their own mobile
phones. For primary school children, smartwatches are distributed to the children, which
are paired to their parents' smartphones.

As a result of these procedures, schools can be selected for participation in the data
collection by the researchers (based on the geographical region of interest), however
students of these schools are free to participate or not (given permission by their parent
or legal guardian, where necessary). This selection process is expected to introduce some
self-selection bias, since the distribution of the participating sample is, in the general
case, different than that of the population of the participating schools. Coverage bias is
also introduced due to children not owning a smartphone, or due to their parents not
owning a smartphone (in the case of primary school children).

Resolving these issues remains an open issue in BigO and the domain of childhood
obesity. Current research is involved with performing appropriate participant segmentation
by clustering behavioral profiles \ref{sec:behavioural_profiles} and use this to define a
sample weighting scheme to counter the effect of these biases.

\section{Aggregation and privacy control mechanisms}
\label{sec:aggregation}

Behavioral information that is described by the indicators of Section
\ref{sec:behaviour_indicators} represents the behavior at individual level. In the
context of evidence-based policy making, however, we are interested in the behavior of the
population at a certain \emph{geographical region} during an \emph{observation period},
which is computed using some type of \emph{aggregation function}.

\begin{figure}
  \includegraphics[width=.95\linewidth]{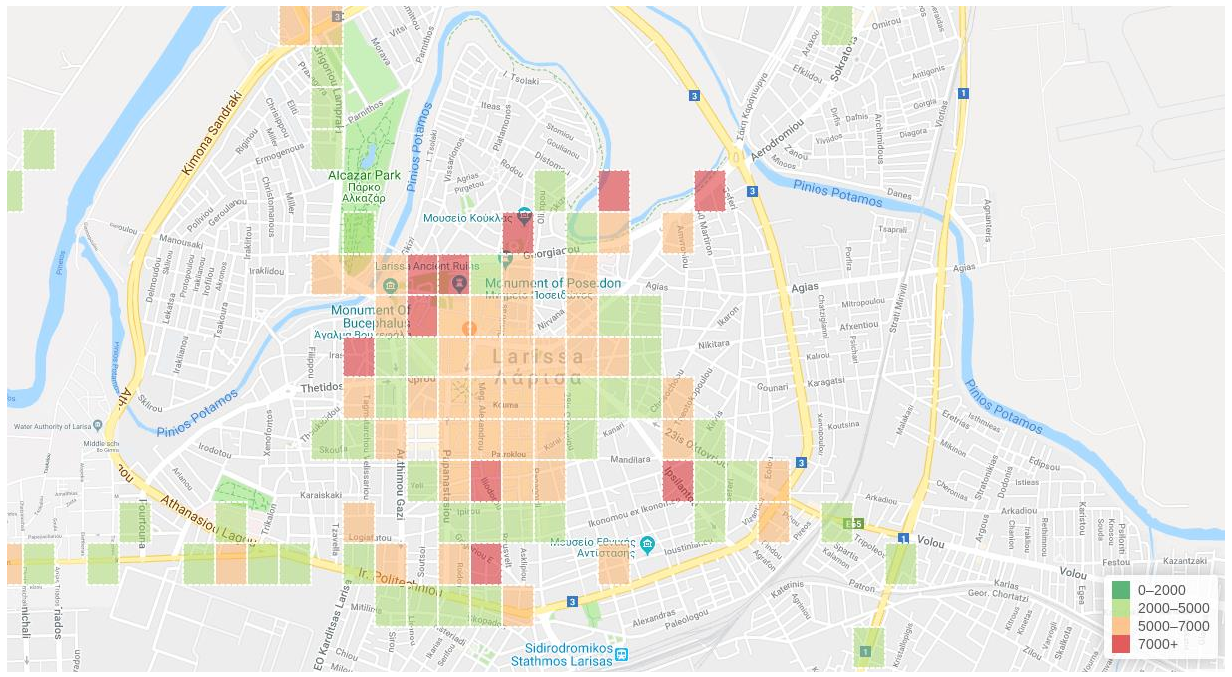}
  \caption{Example choropleth map depicting the geospatial distribution (at 7-character
    geohash level) of the average value of ``activity counts per minute'' in the city
    center at Larissa, Greece. The block of higher (orange) values at the city center can be
    explained by the pedestrian zone that has been built there.}
  \label{fig:larissa}
\end{figure}

Geographical regions of interest can be census units (usually consisting of a few thousand
people each) or larger administrative regions, such as municipalities. For the problem of
childhood obesity we are interested in modeling the local context with high detail and are
therefore interested in high geographical resolution. We have adopted the encoding of geohashes
\cite{geohash} for measuring LECs and aggregating population behaviors. An example showing
aggregated physical activity data at 7-character geohash level is shown in Fig.
\ref{fig:larissa}. This system is similar to the GEOSTAT 2011 population grid
\cite{GEOSTAT} or any other grid-based system which integrates geographical and
statistical data. The geohash encoding has the additional advantage of being able
to easily support different geographical resolutions, since the geographical resolution
depends on the length of the geohash string, with longer geohashes corresponding to
smaller regions.

Regarding the length of observation time, it should be of sufficient duration to capture
the desired behaviors. As a tradeoff between measurement quality and possible burden
placed on participanting children, we consider two weeks of monitoring data as sufficient
for the purposes of measuring obesogenic behaviors. More data should be obtained however,
whenever possible.

In terms of data aggregation, there are two categories of aggregations in the proposed
methodology. Those that aggregate the behavior of the individuals that \emph{live} at a
specific region, and those that aggregate the behavior of individuals \emph{visiting} a
specific region. We can briefly refer to the first type of analysis as the ``habits'' of
the population and the second type as the ``use of resources.'' Not all indicators are
applicable to both types of analysis. For example, the ``Average weekly visits to fast
food restaurants'' cannot be applied to the ``use of resources'' type of analysis, since
it examines the behavior of an individual irrespective of their location. On the
other hand, the ``Number of fast food meals during a visit to a region'' indicator applies
to the ``use of resources'' type of analysis only, since it focuses on the behavior of
people visiting a particular region. Despite their differences, both types of analysis use
similar aggregation functions, which are described in the following section.

\subsection{Aggregation functions}
\label{sec:aggfunctions}

One can think of simple averaging as the most common aggregation function. In more general
(and formal) terms, an aggregation function is a mapping from a set of tuples to a summary
which is most commonly a real number or a distribution (a vector of real numbers that sum
to one),
\begin{equation}
  f: D_1\times D_2\times \dotsc\times D_n \rightarrow R_{agg}
\end{equation}
where $D_k$ is the domain of the $k$-th tuple element, $k = 1,\dotsc,n$ and $R_{agg}$ is
the range of the aggregation function. In our case the tuples are of the form $(U, G, T,
B_i)$ where random variable $G$, corresponds to the geographical region, $T$ to the time
range, $B_i$ to the value of the $i$-th indicator, while $g$ and $b_i$ correspond to specific
values in the domain of these random variables. Random variable $U$
corresponds to an individual from the population. For example, a tuple describing the
number of steps can have the form
\begin{equation}
  \verb|(5, sx0r4k, 20190701T11:52, 32)|
\end{equation}
which is translated as ``individual with id 5 that lives in geohash sx0r4k performed 32
steps at the minute 11:52 of the 1st of July, 2019''\footnote{The corresponding interpretation for the
  ``use of resources'' analysis would indicate that the individual visited geohash sx0r4k
  during that time, and not that he or she lives there.}.

Given this notation, Table \ref{tab:aggfunctions} lists a set of aggregation functions
which are useful for the purposes of collecting evidence for obesogenic behaviors in a
region. Depending on the type of analysis, the expectation may be applied for the
individuals living in the region, or the individuals only during their visit to a region,
as explained previously. It is also worth mentioning that the aggregation functions of
Table \ref{tab:aggfunctions} can also be applied with additional filtering criteria, e.g.,
based on age, gender, value of other indicators etc.

\begin{table}
  \caption{A set of common aggregation functions}
  \label{tab:aggfunctions}
  \begin{tabular}{|l|p{.9\textwidth}|}
    \hline
    \multirow{3}{*}{$f_1$}
    & \textbf{Description:} Average value over individuals in the region\\
    & \textbf{Definition:} Let $\bar{B}_{il}= E\left\{B_i|G=g, U=U_l\right\}$ be the
    average value, over time, of the indicator $B_i$ for user $U_l$ who lives in geohash
    $g$. Then, $f_1(b_i, g) = E\left\{\bar{B}_{il}\right\}$.  \\ 
    & \textbf{Example:} For $B_i$ the number of visits to fast food restaurants for people
    living $g$ during a specific
    week (each tuple corresponds to one week), this aggregation gives the average number
    of weekly visits to fast food restaurants for residents of $g$. \\ 
    \hline
    \multirow{3}{*}{$f_2$}
      & \textbf{Description:} Weighted average, depending on contributed data. This is
    mostly useful for the ``use of resources'' type of analysis \\
      & \textbf{Definition:} $f_2(b_i, g) = E\left\{B_i|G=g\right\}$ \\
      & \textbf{Example:} For $B_i$ the steps per minute walked by individuals
      visiting $g$ (each tuple corresponds to one minute), this aggregation provides the
      average steps per minute across time spent in $g$.\\
    \hline
    \multirow{3}{*}{$f_3$}
    & \textbf{Description:} Probability mass function of the indicator values (if the variable is
    continuous, then its values are grouped into bins) \\
    & \textbf{Definition:} $f_3(b_i, g) = \left[Pr(\bar{B}_{il}=b_{i0}|G=g), \dotsc,
      Pr(\bar{B}_{il}=b_{iM-1}|G=g)\right]^T$,  where $Pr$ is probability and $b_{m}$,
    $m=0,\dotsc,M-1$ are the values (for categorical variables) or the bins (for
    continuous variables). \\
    & \textbf{Example:}  For $B_i$ the transportation mode used at each minute during a
    trip (each tuple corresponds to one minute and only during transportation), this indicator
    provides the distribution of the means of transportation used by residents of $g$
    during their trips.\\ 
    \hline
    \multirow{3}{*}{$f_4$}
      & \textbf{Description:} Percentage of individuals with average $B_i$ under a threshold $t$, for a region \\
      & \textbf{Definition:} $f_4(b_i, g, t)=Pr(\bar{B}_{il} \leq t)$. \\
      & \textbf{Example:} For $B_i$ an indicator of daily steps of an individual (each
    tuple corresponds to a day) and $t = 5000$, this aggregation provides the percentage
    of population that walk, on average, less than 5000 steps per day.\\ 
    \hline
  \end{tabular}
\end{table}

\subsection{Privacy protection}

Privacy protection in the proposed methodology aims at eliminating the possibility of
inferring information about individuals through the aggregated behavior data, as well as
on limiting the sensitive individual information that is stored centrally in the
system. Detailed analysis of the privacy protection mechanisms is beyond the scope of this
paper, however they are briefly outlined here for completeness. They include the following
measures:
\begin{enumerate}
  \item No directly identifiable information (names, emails) is stored. Participation is
    performed through registration codes
  \item Data about individuals is never displayed or shared
  \item Geographical region size is adjusted dynamically to include data above a minimum number of
  individuals
\item Support for distributed computation
\end{enumerate}

The first two mechanisms are enforced by design. Individual data are stored in a different
database than the aggregated data. Data from the database containing sensitive, individual
data is never shared or used for display.

Even when providing only aggregated data, however, there are privacy risks when the number
of individuals is small. Assuming that only few participants (i.e. below a threshold) have
provided data for a region, it is possible to (a) disregard the region or (b) use larger
regions, until sufficient participants are included. This second approach has the
advantage that valuable data is not ignored in the analysis. In the case of geohashes,
this can be achieved by reducing the geohash length until the required number of
participants is included.

Finally, distributed computation protects highly sensitive individual data (such as raw
location data), by analyzing them at the edge device, without transmitting them and
storing them centrally. According to this approach, the raw sensor data is processed at
the participants' smartphones to extract the behavioral indicators of Section
\ref{sec:behaviour_indicators}. The indicators are then transmitted and stored for
analysis. The drawback of this approach is that the raw data is not available later, if
additional processing needs to be done.

\section{Quantifying data quality}
\label{sec:quality}

In contrast to data collection under controlled conditions, working with big
data sources introduces significant data quality challenges. Low data
quality in the context of this work can be due to
\begin{itemize}
  \item \emph{Missing or incomplete data}. E.g., some participants will only wear a
    smartwatch a few hours of the day or a few days per week, while others will deactivate
    location data recording.
  \item \emph{Heterogeneous data sources}. E.g., the accuracy of physical activity
    indicators depends on the type of accelerometer sensor and its sampling rate.
  \item \emph{Measurement errors for behavioral indicators}. E.g., activity
    type recognition or transportation mode detection algorithms are not 100\% accurate.
\item \emph{Bias}. E.g., selection bias (differences between those who choose to participate and
    those who don't), coverage bias (differences between those who are able to
    participate and those who don't).
\end{itemize}

Regarding the last error type (bias), work on determining appropriate sample weighting to
counter the effect of selection and coverage bias is currently in progress (as discussed
in Section \ref{sec:bias}) and is not discussed here. For the first three sources of
error, our approach is to quantify the quality of each data sample using a common, 5-level
data quality scale, which is shown in Table \ref{tab:quality}
\begin{table}
  \caption{Data quality levels}
  \label{tab:quality}
  \begin{tabular}{|p{.1\linewidth}|p{.05\linewidth}|p{.7\linewidth}|}
    \hline
    \textbf{Quality} & \textbf{Value} & \textbf{Example} \\
    \hline
    Very low & 0.2 & Aggregation from a small number of participants, as determined by the
    variance of the indicator value \\
    Low & 0.4 & Estimating physical activity from a mobile phone that is carried for less
    than 2 hours per day\\
    Moderate & 0.6 & Use of surrogate variables to estimate LECs\\
    High & 0.8 & Identification of visited POI types based on location data and external
    databases\\
    Very High & 1.0 & LECs provided by official statistics\\
    \hline
  \end{tabular}
\end{table}

Having a quantified quality level for each measurement allows us to represent our
``confidence'' in the measurement and take this information into account in subsequent
statistical analysis \cite{Kutner2005} or predictive modeling \cite{Sarafis2018} steps. In
the following subsections we briefly outline a set of simple guidelines for quantifying
data quality.

\subsection{Quality determined by data availability}

Missing values and incomplete data is a result of measuring using general-purpose
wearables and sensors. Not all devices include all sensors, while compliance across users
varies. In addition, behavior measurements for each geographical region can be computed
from different numbers of users, with different recording duration.

To map data availability to quality levels, we use thresholds. A ``Very low'' threshold
determines the value under which the quality is ``Very low'', while a ``Very high''
threshold is used to determine the value above which the quality is ``Very high.'' Values
in-between are linearly interpolated to determine the quality level. The thresholds
presented in Table \ref{tab:availability} are indicative and depend on the application and
type of behavior that needs to be measured.
\begin{table}
  \caption{Mapping data availability to quality level}
  \label{tab:availability}
  \begin{tabular}{|p{0.25\linewidth}|p{0.1\linewidth}|p{0.1\linewidth}|p{0.4\linewidth}|}
    \hline
    \textbf{Data type} & \textbf{Very low threshold} & \textbf{Very high threshold} &
    \textbf{Comments} \\
    \hline
    Daily duration of accelerometer recordings & 1 hour & 6 hours & Occurs because the
    device is not used, or because data acquisition process is stopped by the operating
    system\\
    Daily duration of GPS recordings & 1 hour & 6 hours & Same as accelerometer. Also, users
    have the option of turning off GPS\\
    Data recorded per region & 10 hours & 100 hours & Values are indicative. Quality
    depends on the region size and recording variance\\
    Number of users providing data per region & 10 & 100 & As above, actual values depend
    on the desired statistical power\\
    \hline
  \end{tabular}
\end{table}

\subsection{Quality determined by data source and accuracy of behavioral indicator extraction}

When measuring behaviors for the purpose of understanding and preventing childhood obesity
we consider the quality of data produced by smartwatches to be ``Very high.'' For data
produced by mobile phones, the quality is ``High.'' Regarding LECs, data coming from
statistical authorities are considered to be of ``Very high'' quality. The quality of map
and GIS data sources for estimating LECs varies. For example our experience is that
Google's maps (``Moderate'' quality) are less reliable than Foursquare maps (``Very high''
quality) regarding available venues. Depending on the application, a small number of
experiments can allow assignment of quality levels to different data sources.

Measurement errors for behavioral indicators are introduced by the behavioral indicator
extraction algorithms. Based on the algorithm effectiveness, as measured in annotated
datasets, one can estimate the quality level for each indicator. A discussion on the
effectiveness of the various indicator extraction algorithms is, however, beyond the scope
of this work.

\subsection{Multiple simultaneous sources of error}

Multiple sources of error may be present simultaneously during the analysis. For example,
the estimated average number of daily steps a child performs in a region may be inaccurate
due to low sample size, due to missing measurements during the day, or because the step
counting algorithm introduces error. To keep the analysis simple, we treat the quality
levels as fuzzy numbers and use fuzzy operators to combine them, such as fuzzy
intersection and union ($t$-norms and $t$-conorms, respectively) \cite{Klir1995}. For
example, if the data quality levels determined by data availability and behavioral
indicator extraction are $m_1$ and $m_2$ respectively, then we can determine an overall
sample quality level using the standard fuzzy intersection (the minimum of the values) as
\begin{equation}
  m_{12} = \min(m_1, m_2)
\end{equation}
On the other hand, if we have two measurement types for the same information (e.g.,
self-reports and objectively measured information on fast food visits) then we can expect
the quality of our data to be the union of the two values. Using the standard union,
\begin{equation}
  m_{12} = \max(m_1, m_2)
\end{equation}
The advantage of this approach is that any type of $t$-norm and $t$-conorm can be used,
depending on the needs of each application. The reader is referred to \cite{Klir1995} for
a list of the most commonly used fuzzy intersection and union operators. As in the
previous, the overall quality level can then be used as a weight or a fuzzy number in the
statistical analysis or predictive modeling procedures.

\section{Challenges, open issues and lessons learned}
\label{sec:lessons_learned}

This section provides an informal account of the obstacles that we have encountered, thus
far, while implementing the proposed methodology in the BigO technology platform. Given the
overlap between BigO and Trusted Smart Surveys (as discussed Section
\ref{sec:introduction}), our hope is that such information will be useful for those who
develop similar solutions for Trusted Smart Statistics.

\subsection{Technology}

The technical challenges that we encountered are the result of the requirements for sensor
data acquisition from commodity smartphones and smartwatches. These requirements include
the following:
\begin{enumerate}
  \item Data should originate from off-the-shelf smartphones or smartwatches that
    participants already own (using special-purpose devices is not a viable option for
    extracting population-level statistics)
  \item Data collection should be as unobtrusive as possible, to increase usability and
    compliance. Aside from answering questions (active data collection) users should not
    notice any changes in their device's operation while sensor data are passively
    collected
  \item Most of the data processing should take place at the edge device, to avoid
    unnecessary transmission of personal data
  \item Behavioral indicators should be as accurate as possible. In any case, the
    probability of error should be quantified
\end{enumerate}

The main problem associated with the first two requirements is that the operating systems
of modern smartphones and smartwatches (namely Android, Android Wear OS and iOS) have
built-in mechanisms for battery saving which prevent applications to execute continuously
in the background. This means that unless special provisions are made from the developer
side (such as the ones outlined in Section \ref{sec:doze}), data acquisition may
unexpectedly stop when the data acquisition application runs in the background. To make
matters worse, many vendors of popular Android devices have introduced special,
non-standard and non-documented procedures for stopping applications in the
background. These cannot be bypassed programmatically, so device-specific multi-step
instructions need to be provided to the users of such smartphones.

For the third requirement (processing at the edge device), complex processing can indeed
be carried out in modern smartphones, since they are equipped with powerful processing
units. It is best to collect the data locally and perform the processing when the device
is charging, since in that case any processing side-effects (power consumption, increased
device temperature) are not noticable to the users. On the other hand, smartwatches are
less appropriate for data processing and a mechanism must be implemented for transmitting
the data to the paired smartphone first.

Finally, regarding the fourth requirement, research on signal processing and machine
learning algorithms for behavioral indicator extraction is still in progress, although the
state of the art is already fairly accurate \cite{Brajdic2013}, \cite{Kyritsis2019},
\cite{Luo2017}, \cite{Papapanagiotou2018}, \cite{Reiss2012}, \cite{Wang2018}. One problem
is that development and evaluation of these algorithms take place using publicly available
datasets which generally are different compared to the data collected through mobile and
wearable applications. Additional algorithm development and validation is therefore
needed. In the same context, it is important to highlight that errors are also introduced
by limitations on the use of the system. For example, people often don't carry their
mobile phones when exercising. This means that often the individual physical activity is
understimated.

As a result of all the above issues, is important to plan for significantly increased
research and application development time compared to other mobile applications which
usually focus on front-end development. This also depends on the required sensors (IMU
sensors are usually more complex to handle than GPS, for example). Given that data
acquisition is a prerequisite for all subsequent data processing and analysis steps, an
overly optimistic development time estimate here will have a major impact to project
planning.

\subsection{Usability and participation}

Usability is cruicially important for citizen science applications, since it facilitates
participation, inclusion and retention (especially when longer duration of use is
desired). Our experience so far is that existing devices, and especially smartwatches, are
not sufficiently user-friendly for many users who expect seamless, minimum-effort setup
procedures. To overcome this barrier, special emphasis must be placed on user experience
and interaction, and especially on making sure that adequate feedback is provided to
users. This includes feedback showing whether data is being recorded or not, as well as
appropriate indications when something is wrong and what the user should do about it, both
for the setup and the normal operation of the system.

In addition to usability, voluntary participation requires the use of some type of
incentive and an engagement mechanism. For the children who are citizen-scientists of
BigO, user engagement is mainly achieved through school-based activities, coordinated by
teachers. These can be quite effective for children, but do not generalize to the entire
population and require effort from a third-party (i.e., the teachers in our
case). Research, experimentation and possibly several pilot deployment rounds are required
to discover effective incentives and communication strategies that work in engaging the
required population sample in the case of Trusted Smart Surveys.

\section{Conclusions}
\label{sec:conclusions}

We have presented an overview of the BigO methodology for collecting evidence on
population behavior and the environment related to the problem of childhood obesity. BigO
develops tools that allow for the monitoring of obesogenic behaviors of the population and
the association of these behaviors with the characteristics of the
environment. Individuals voluntarily offer their data according to the citizen scientist
paradigm. The personal sensory data originating from worn IMU sensors, GPS, pictures
captured by the users and responses to questionnaires are being aggregated in order to
produce behavioral indicators and behavioral profiles. Collection of data from open and
online sources is leveraged to produce LECs, which refer to geo-aligned POIs and
statistical variables (demographic, social, financial, etc.) known to be linked to obesity
and unhealthy lifestyle choices. Extracted behavioral indicators are being correlated to
LECs as a means to identify local factors that cause (childhood) obesity.
Furthermore, BigO adopts strict privacy preservation mechanisms, including innovative
aggregation methods, and features data quality criteria that take into account various
sources of error.

\section*{Acknowledgements}

The work leading to these results has received funding from the European Community's
Health, demographic change and well-being Programme under Grant Agreement No. 727688
\protect\url{(http://bigoprogram.eu)}, 01/12/2016 - 30/11/2020.

\bibliographystyle{vancouver}
\bibliography{biblio}

\begin{thebibliography}{10}

\bibitem{Bammann2013}
Bammann K, Gwozdz W, Lanfer A, Barba G, De~Henauw S, Eiben G, et~al.
\newblock Socioeconomic factors and childhood overweight in Europe: results
  from the multi-centre IDEFICS study.
\newblock Pediatric obesity. 2013;8(1):1--12.

\bibitem{Wijnhoven2014}
Wijnhoven TM, van Raaij JM, Spinelli A, Starc G, Hassapidou M, Spiroski I,
  et~al.
\newblock WHO European Childhood Obesity Surveillance Initiative: body mass
  index and level of overweight among 6--9-year-old children from school year
  2007/2008 to school year 2009/2010.
\newblock BMC public health. 2014;14(1):806.

\bibitem{EUActionPlan2014}
EU Action Plan on Childhood Obesity 2014-2020;.
\newblock Available online:
  \url{https://ec.europa.eu/health/sites/health/files/nutrition_physical_activity/docs/childhoodobesity_actionplan_2014_2020_en.pdf}.

\bibitem{WHOReport}
Organization WH. Report of the commission on enbding childhood obesity; 2016.
\newblock Available online:
  https://www.who.int/end-childhood-obesity/publications/echo-report/en/.

\bibitem{Lyn2013}
Lyn R, Aytur S, Davis TA, Eyler AA, Evenson KR, Chriqui JF, et~al.
\newblock Policy, systems, and environmental approaches for obesity prevention:
  a framework to inform local and state action.
\newblock Journal of public health management and practice: JPHMP. 2013;19(3
  Suppl 1):S23.

\bibitem{Lobstein2015}
Lobstein T, Jackson-Leach R, Moodie ML, Hall KD, Gortmaker SL, Swinburn BA,
  et~al.
\newblock Child and adolescent obesity: part of a bigger picture.
\newblock The Lancet. 2015;385(9986):2510--2520.

\bibitem{Wang2013}
Wang Y, Wu Y, Wilson RF, Bleich S, Cheskin L, Weston C, et~al.
\newblock Childhood obesity prevention programs: comparative effectiveness
  review and meta-analysis.
\newblock In: Database of Abstracts of Reviews of Effects (DARE):
  Quality-assessed Reviews [Internet]. Centre for Reviews and Dissemination
  (UK); 2013. .

\bibitem{DeBour2015}
De~Bourdeaudhuij I, Verbestel V, De~Henauw S, Maes L, Huybrechts I, M{\aa}rild
  S, et~al.
\newblock Behavioural effects of a community-oriented setting-based
  intervention for prevention of childhood obesity in eight European countries.
  Main results from the IDEFICS study.
\newblock Obesity reviews. 2015;16:30--40.

\bibitem{bigosite}
BigO: Big data against childhood obesity. H2020 Research project, Grant
  Agreement No: 727688; 2019.
\newblock Available online: \url{http://bigoprogram.eu}.

\bibitem{Ricciato2019}
Ricciato F, Wirthmann A.
\newblock Trusted Smart Statistics: {H}ow new data will change official
  statistics.
\newblock In: 4th International Conference on Data for Policy; 2019. .

\bibitem{Archer2018}
Archer DW, Bogdanov D, Lindell Y, Kamm L, Nielsen K, Pagter JI, et~al.
\newblock From Keys to Databases—Real-World Applications of Secure
  Multi-Party Computation.
\newblock The Computer Journal. 2018;61(12):1749--1771.

\bibitem{Zyskind2015}
Zyskind G, Nathan O, Pentland A.
\newblock Enigma: Decentralized computation platform with guaranteed privacy.
\newblock arXiv preprint arXiv:150603471. 2015;.

\bibitem{BucharestMemorandum2018}
Bucharest memorandum on official statistics in a datafied society (trusted
  smart statistics); 2018.
\newblock Available online:
  \url{http://www.dgins2018.ro/bucharest-memorandum/}.

\bibitem{Kyritsis2019}
{Kyritsis} K, {Diou} C, {Delopoulos} A.
\newblock Modeling Wrist Micromovements to Measure In-Meal Eating Behavior from
  Inertial Sensor Data.
\newblock IEEE Journal of Biomedical and Health Informatics. 2019;.

\bibitem{Kyritsis2018}
{Kyritsis} K, {Diou} C, {Delopoulos} A.
\newblock End-to-end Learning for Measuring in-meal Eating Behavior from a
  Smartwatch.
\newblock In: 2018 40th Annual International Conference of the IEEE Engineering
  in Medicine and Biology Society (EMBC); 2018. p. 5511--5514.

\bibitem{Papadopoulos2018}
{Papadopoulos} A, {Kyritsis} K, {Sarafis} I, {Delopoulos} A.
\newblock Personalised meal eating behaviour analysis via semi-supervised
  learning.
\newblock In: 2018 40th Annual International Conference of the IEEE Engineering
  in Medicine and Biology Society (EMBC); 2018. p. 4768--4771.

\bibitem{Doze}
{Android OS Developers Documentation}. Optimize for Doze and App Standby;.
\newblock Available online:
  \url{https://developer.android.com/training/monitoring-device-state/doze-standby}.

\bibitem{COSI}
WHO European Childhood Obesity Surveillance Initiative (COSI);.
\newblock Available online:
  \url{http://www.euro.who.int/en/health-topics/disease-prevention/nutrition/activities/who-european-childhood-obesity-surveillance-initiative-cosi}.

\bibitem{Diou2018}
Diou C, Lelekas P, Delopoulos A.
\newblock Image-Based Surrogates of Socio-Economic Status in Urban
  Neighborhoods Using Deep Multiple Instance Learning.
\newblock Journal of Imaging. 2018;4(11):125.

\bibitem{Tryon1996}
Tryon WW, Williams R.
\newblock Fully proportional actigraphy: A new instrument.
\newblock Behavior Research Methods, Instruments, {\&} Computers. 1996
  Sep;28(3):392--403.
\newblock Available from: \url{https://doi.org/10.3758/BF03200519}.

\bibitem{Sarafis2019}
{Sarafis} I, {Diou} C, {Delopoulos} A.
\newblock Behaviour Profiles for Evidence-based Policies Against Obesity.
\newblock In: 2019 41th Annual International Conference of the IEEE Engineering
  in Medicine and Biology Society (EMBC); 2019. .

\bibitem{Luo2017}
Luo T, Zheng X, Xu G, Fu K, Ren W.
\newblock An Improved DBSCAN Algorithm to Detect Stops in Individual
  Trajectories.
\newblock ISPRS International Journal of Geo-Information. 2017;6(3).
\newblock Available from: \url{https://www.mdpi.com/2220-9964/6/3/63}.

\bibitem{Jahromi2016}
Jahromi KK, Zignani M, Gaito S, Rossi GP.
\newblock Simulating human mobility patterns in urban areas.
\newblock Simulation Modelling Practice and Theory. 2016;62:137 -- 156.
\newblock Available from:
  \url{http://www.sciencedirect.com/science/article/pii/S1569190X15001732}.

\bibitem{geohash}
Geohash official site;.
\newblock Available online: \url{http://geohash.org}.

\bibitem{GEOSTAT}
Eurostat. Statistics Explained: Population grids;.
\newblock Available online:
  \url{https://ec.europa.eu/eurostat/statistics-explained/index.php/Population_grids#Grid_statistics}.

\bibitem{Kutner2005}
Kutner MH, Nachtsheim CJ, Neter J, Li W, et~al.
\newblock Applied linear statistical models.
\newblock McGraw-Hill Irwin Boston; 2005.

\bibitem{Sarafis2018}
Sarafis I, Diou C, Delopoulos A.
\newblock Span error bound for weighted SVM with applications in hyperparameter
  selection.
\newblock arXiv preprint arXiv:180906124. 2018;.

\bibitem{Klir1995}
Klir GJ, Yuan B.
\newblock Fuzzy sets and fuzzy logic: theory and applications.
\newblock Prentice Hall; 1995.

\bibitem{Brajdic2013}
Brajdic A, Harle R.
\newblock Walk detection and step counting on unconstrained smartphones.
\newblock In: Proceedings of the 2013 ACM international joint conference on
  Pervasive and ubiquitous computing. ACM; 2013. p. 225--234.

\bibitem{Papapanagiotou2018}
Papapanagiotou V, Diou C, Ioakimidis I, S{\"o}dersten P, Delopoulos A.
\newblock Automatic analysis of food intake and meal microstructure based on
  continuous weight measurements.
\newblock IEEE journal of biomedical and health informatics.
  2018;23(2):893--902.

\bibitem{Reiss2012}
Reiss A, Stricker D.
\newblock Creating and benchmarking a new dataset for physical activity
  monitoring.
\newblock In: Proceedings of the 5th International Conference on PErvasive
  Technologies Related to Assistive Environments. ACM; 2012. p.~40.

\bibitem{Wang2018}
Wang L, Gjoreskia H, Murao K, Okita T, Roggen D.
\newblock Summary of the sussex-huawei locomotion-transportation recognition
  challenge.
\newblock In: Proceedings of the 2018 ACM International Joint Conference and
  2018 International Symposium on Pervasive and Ubiquitous Computing and
  Wearable Computers. ACM; 2018. p. 1521--1530.

\end{thebibliography}

\end{document}